\documentclass{PoS}
\usepackage{graphicx}

\def\beq{\begin{equation}}
\def\eeq{\end{equation}}
\def\bea{\begin{eqnarray}}
\def\eea{\end{eqnarray}}
\def\tr{\mathrm{tr}}

\def\ket#1{\left|#1\right>}

\title{Topological Charge in Two Flavors QCD with Optimal Domain-Wall Fermion}

\ShortTitle{Topological Charge in Lattice QCD with ODWF}

\author{%
    \speaker{Tung-Han~Hsieh}$^1$, 
    Ting-Wai~Chiu$^{2,3}$, 
    Yao-Yuan~Mao$^2$
    (for the TWQCD Collaboration) \\
    $^1$ Research Center for Applied Sciences, Academia Sinica, Taipei 115, Taiwan \\
    $^2$ Department of Physics, and Center for Theoretical Sciences, National Taiwan University, Taipei 10617, Taiwan \\
    $^3$ Center for Quantum Science and Engineering, National Taiwan University, Taipei 10617, Taiwan \\
}

\abstract{%
We measure the topological charge
of the gauge configurations generated by lattice simulations
of 2 flavors QCD on a $ 16^3 \times 32 $ lattice,  
with Optimal Domain-Wall Fermion (ODWF) at $ N_s = 16 $  
and plaquette gauge action at $ \beta = 5.90 $.  
Using the Adaptive Thick-Restart Lanczos algorithm, 
we project the low-lying modes of the 4D effective Dirac operator of ODWF,  
and obtain the topological charge and topological susceptibility.
Our result of topological susceptibility agrees with the sea-quark
mass dependence predicted by the chiral perturbation theory, 
and provides a determination of the chiral condensate.}

\FullConference{The XXVIII International Symposium on Lattice Field Theory, Lattice 2010, 
		June 14-19, 2010, 
		Villasimius, Italy}

\begin{document}

\section{Introduction}

In Quantum Chromodynamics (QCD), the topological susceptibility ($\chi_t$)
is the most essential quantity to measure the topological charge
fluctuation of the QCD vacuum, which plays an important role in breaking
the $U_A(1)$ symmetry. Theoretically, $\chi_t$ is defined as
\beq
\label{eq:chi_t}
\chi_{t} = \int d^4 x  \left< \rho(x) \rho(0) \right>
         = \frac{\left< Q_t^2\right>}{\Omega}
\eeq
where
\beq\label{eq:topFuv}
\rho(x) = \frac{1}{32 \pi^2} \epsilon_{\mu\nu\lambda\sigma}
                             \tr[ F_{\mu\nu}(x) F_{\lambda\sigma}(x) ],
\hspace{5mm}
Q_t \equiv \int d^4 x \rho(x),
\eeq
$Q_t$ is the topological charge (which is an integer for QCD),
$\rho(x)$ is the topological charge density, and $\Omega$ is the space-time
volume. In the chiral perturbation theory (ChPT), the quark mass 
dependence of $\chi_t$ was derived at the tree level \cite{Leutwyler:1992yt} 
in 1992, and it has been extended to the one-loop order recently \cite{Mao:2009sy}. 
For $ N_f = 2 $, these ChPT formulas can be written as  
\bea\label{eq:chi_ChPT}
\chi_t/m_u &=& \frac{\Sigma}{1 + m_u/m_d}, \\
\label{eq:chi_ChPT_NLO}
\chi_t/m_u &=& \frac{\Sigma}{1 + m_u/m_d}\left[
        1 - \biggl(\frac{3 M_{\pi}^2}{32\pi^2 F_{\pi}^2}\biggr)
        \ln\frac{M_{\pi}^2}{\mu^2_{sub}}
	+K_6(m_u+m_d) + 2(2 K_7 + K_8)\frac{m_u m_d}{m_u+m_d}\right], 
\eea
where $\Sigma$ is the chiral condensate, $ F_\pi $ the pion decay constant, 
and $ K_i $  are related to low energy constants $ L_i $.
A salient feature of $ \chi_t $ is that it is suppressed in
the chiral limit ($m_u \rightarrow 0$) due to the internal quark loops. 
Most importantly, (\ref{eq:chi_ChPT}) and (\ref{eq:chi_ChPT_NLO}) provide a viable
way to extract $\Sigma$ and other low energy constants.

However, one cannot determine $Q_t$ reliably using the link variables, 
due to the rather strong fluctuations at short distances. 
Instead, we consider the Atiyah-Singer index theorem
\cite{Atiyah:1968mp}
\beq\label{eq:AS_thm}
Q_t = n_+ - n_- = \mbox{index}({\cal D}), 
\eeq
where $ n_\pm $ is the number of zero modes of the massless Dirac
operator $ {\cal D} \equiv \gamma_\mu ( \partial_\mu + i g A_\mu) $
with $ \pm $ chirality. 
Thus, by couting the zero modes of the massless Dirac operator,  
we can determine $Q_t$ and $ \chi_t $ for an ensemble of gauge configurations. 

Recently, the topological susceptibility  
has been measured in unquenched lattice QCD with exact chiral symmetry, 
for $ N_f = 2 $ lattice QCD with overlap fermion 
in a fixed topology \cite{Aoki:2007pw},  
and $ N_f = 2+1 $ lattice QCD with domain-wall fermion \cite{Chiu:2008jq}. 
These results are in good agreement with the ChPT at the tree level (\ref{eq:chi_ChPT}).

In this work, we measure the topological charge 
of the gauge configurations generated by lattice simulations
of 2 flavors QCD on a $ 16^3 \times 32 $ lattice, 
with the Optimal Domain-Wall Fermion (ODWF) \cite{Chiu:2002ir} at $ N_s = 16 $, 
and plaquette gauge action at $ \beta = 5.90 $.  
Mathematically, ODWF is a theoretical framework which can preserve 
the chiral symmetry optimally for any finite $N_s$. 
Thus the artifacts due to the chiral
symmetry breaking with finite $ N_s $ can be reduced to the minimum, 
especially in the chiral regime.
The 4D effective operator of massless ODWF is
\beq\label{eq:ov}
D = m_0 [1+ \gamma_5 S_{opt}(H_w) ], \quad 
S_{opt}(H_w) = \frac{1-\prod_{s=1}^{N_s} T_s}{1 + \prod_{s=1}^{N_s} T_s}, \quad 
T_s = \frac{1-\omega_s H_w}{1+\omega_s H_w}, 
\eeq
which is exactly equal to the Zolotarev optimal rational approximation 
of the overlap Dirac operator \cite{Neuberger:1997fp}. That is,   
$ S_{opt}(H_w) = H_w R_Z(H_w) $, where $ R_Z(H_w)$ is the optimal 
rational approximation of $ (H_w^2)^{-1/2} $ \cite{Chiu:2002eh}.

We outline our scheme of projecting the low-lying eigenmodes of $D$ as follows. 
Using $ [D D^{\dagger}, \gamma_5]=0 $, and the representation
\beq
\label{eq:eigen-Dov}
D\ket{\theta} = \lambda(\theta)\ket{\theta},\hspace{10mm}
\lambda(\theta) = m_0 (1+e^{i\theta}), 
\eeq
we obtain
\beq
\label{eq:eigen-Dov-pm}
S_{\pm}\ket{\theta}_{\pm} 
\equiv P_{\pm} H_w R_Z(H_w) P_{\pm} \ket{\theta}_{\pm}
= \pm \cos\theta\ket{\theta}_{\pm}, \quad 
P_{\pm} = \frac{1 \pm \gamma_5}{2},   
\eeq
where $\ket{\theta} = P_+\ket{\theta} + P_-\ket{\theta} =
\ket{\theta}_+ + \ket{\theta}_-$.
Thus, we can perform the eigenmode
projection on the operator $S_{\pm}$ instead of $ D $.
Moreover, $\ket{\theta}_{\pm}$ are related to each other through the relation 
\beq
\label{eq:DovEiv}
\ket{\theta} = \frac{1}{i\sin\theta}(\gamma_5 S -e^{-i\theta})\ket{\theta}_\pm, 
\hspace{10mm}
\theta \ne 0, \pm\pi, \pm 2\pi, \ldots
\eeq
Assuming the zero modes only residing in either $ + $ or $ - $ chirality, 
we project the low-lying modes of $ D $ as follows. 
First, we project the smallest eigenmodes of $S_+$. 
If $D$ has zero modes with positive chirality, then the smallest eigenvalues 
of $ S_+ $ are equal to $-1$, and the corresponding full eigenvectors can be obtained 
using Eq. (\ref{eq:DovEiv}). Otherwise, we have to check whether $ D $ has zero modes 
with negative chirality by projecting the largest eigenmodes of $ S_- $.
If $ D $ has zero modes with negative chirality, the largest eigenvalues of $ S_- $
are equal to $+1$, and the corresponding full eigenvectors can be obtained 
using Eq. (\ref{eq:DovEiv}).
Finally, for non-zero low-lying eigenmodes, we can pick either $ + $ or $ - $ chirality
of Eq. (\ref{eq:eigen-Dov-pm}) for projection, then obtain the full 
eigenvectors using Eq. (\ref{eq:DovEiv}).


\section{The {\em TRLan} algorithm}

In general, the low-mode projection of a large sparse matrix $A$ can
be carried out via iterative process, such as the {\em Lanczos algorithm} or
the {\em Arnoldi algorithm}. The common procedure is to construct an
orthonormal basis from the Krylov subspace starting from an
initial vector $r_0$, 
\beq
\mathcal{K}(A,r_0) = 
{\mathrm span} \left\{ r_0,\, A r_0,\, A^2 r_0, \ldots,\, A^{m-1} r_0 \right\}.
\eeq
The linear combinations of 
$\{ A^i r_0, i=0, 1, \ldots m-1 \}$ form the Ritz vectors
of $A$, which converge to the eigenvectors of $ A $
as $m$ becomes very large. 

\begin{figure}[th]
\begin{tabular}{@{}cc@{}}
\raisebox{23.5mm}{\framebox{%
\begin{minipage}{4.9cm}
\parindent=1.5mm
\vspace{3mm}
{\bf Input}: $r_0$, $\beta_0 = ||r_0||$, $q_0 = 0$
\vspace{1mm}

{\bf For}: $i$ = 1, 2, \ldots
\vspace{-3mm}
\begin{itemize}
\itemsep=0mm
\parskip=0mm
\item $q_i = r_{i-1}/\beta_{i-1}$
\item $p = A q_i$
\item $\alpha_i = q_i^\dagger p$
\item $r_i = p-\alpha_i q_i -\beta_{i-1}q_{i-1}$
\item $\beta_i = ||r_i||$
\end{itemize}
\vspace{3mm}
\end{minipage}}}
&
\includegraphics[width=95mm]{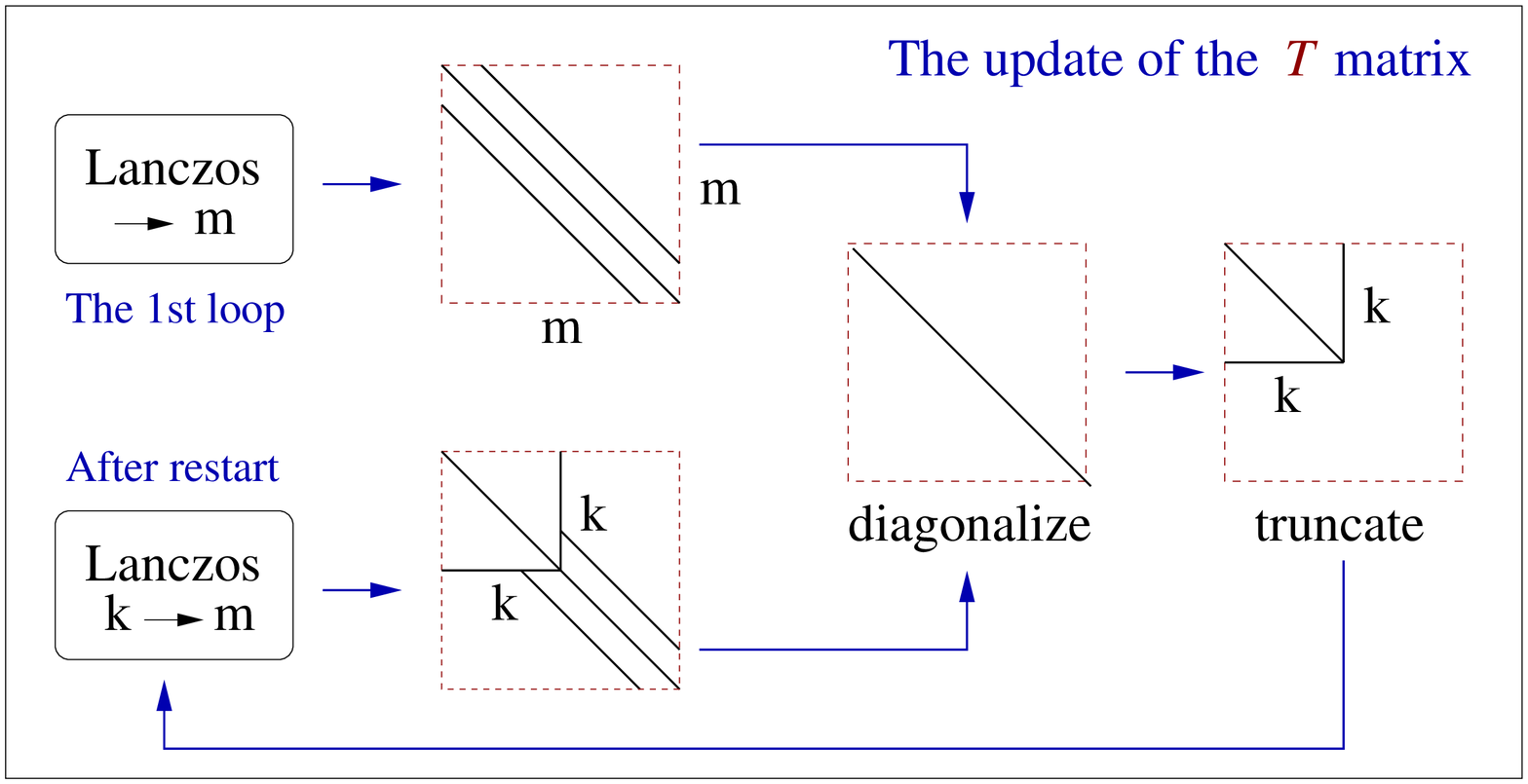} \\
(a) & (b)
\end{tabular}
\caption{\label{fig:TRLan}%
(a) The basic Lanczos process;
(b) The Lanczos algorithm with Thick-Restart.}
\end{figure}

Since $H_w$ and $S_{\pm}$ are Hermitian matrices, it is natural to use the Lanczos algorithm.
The basic Lanczos process (see Fig.\ref{fig:TRLan}-(a)) is to 
construct the following factorization
\beq
\label{eq:lanczFactor}
A Q_m = Q_m T_m + \beta_m q_{m+1} e_m^T, 
\eeq
where $\{ q_1$, $q_2$, \ldots, $q_m \}$ form an orthonormal basis,  
$Q_m=[ q_1, q_2, \ldots, q_m ]$, $ e_m $
is the $m$-th column of the $m \times m $ identity matrix, and
$T_m$ is a tridiagonal matrix with diagonal elements $\alpha_i$ 
and sub-diagonal elements $\beta_i$. 
Then the Ritz pairs $(\hat\lambda_i,\hat x_i)$ can be obtained from
\beq
\hat T_m = U^{\dagger}_m T_m U_m, \hspace{10mm} X_m = Q_m U_m, 
\eeq
where $\hat T_m$ is diagonal with eigenvalues $\hat\lambda_i$, $U_m$ is
a unitary matrix, and $X_m = [\hat x_1, \hat x_2, \ldots, \hat x_m ]$.
As $m$ becomes very large,  
the Ritz pairs $(\hat\lambda_i,\hat x_i)$ converge to the eigenmodes
$(\lambda_i,x_i)$ of $A$.

However, the Lanczos algorithm suffers from the following problems.
Some specious Ritz values may repeatedly appear when $m$ goes to larger values.
This is due to the fact that the vectors $\{ q_i \}$ lose orthogonality 
rapidly in the finite precision arithmatics. Hence one
has to apply the Gram-Schmidt procedure to re-orthogonalize them 
during the iteration.
Also, it requires a large number of $q_i$ to project a few Ritz
pairs, consuming a large amount of memory and 
slowing down the computation. A way to overcome these problems is to
perform a {\em restart}.

There are several restart schemes,
in which the Thick-Restart Lanczos algorithm (TRLan)
\cite{TRLan} turns out to be the most efficient for our purpose.  
The TRLan algorithm is illustrated in Fig.\ref{fig:TRLan}-(b), 
where the squares in the diagram stand for the square matrix $T_m$,
in which the non-zero matrix elements are denoted by black lines. 

\begin{table}[t!]
\begin{tabular}{@{}l|ccc@{\ \ }c|ccc@{\ \ }c@{}}
       & \multicolumn{4}{c}{$H_w$: $k'=240$, $m=340$}
       & \multicolumn{4}{|c}{$S_+$: $k'=100$, $m=200$, $n_p=240$} \\
method &  N\_restarts  & \# of $ A \cdot v $ & time(s) & speed up 
       &  N\_restarts  & \# of $ A \cdot v $ & time(s) & speed up \\\hline
ARPACK         &    388   & 35390    & 136460  & 1.00
               &     13   & 1050     & 112632  & 1.00 \\ 
TRLan          &    999   & 100140   & 572951  & 0.24
               &     12   & 1300     & 105790  & 1.06 \\
$\nu$-TRLan    &    383   & 59145    &  78058  & 1.75
               &     11   & 1030     &  90496  & 1.24
\end{tabular}
\caption{\label{tab:benchmark}
The benchmark of low-mode projection of $H_w$ and $S_+$ for 
a non-trivial gauge configuration ($Q_t = 3$) 
at $\beta=5.9$ and $m_q a = 0.01$. 
Here $k'$ is the number of low-modes projected, $m$ is the
dimension of Krylov subspace, and $n_p$ is the number of low-modes 
of $H_w$ used for low-mode preconditioning 
in the projection of the low-modes of $S_+$.}
\end{table}

Obviously, the performance of the TRLan algorithem depends on the truncated dimension $k$ 
and the dimension $m$ of the Krylov subspace. 
Now the question is what is the optimal value of $k$ (for a given $m$)
such that for each restart, the reduction factor $d_j$ of the residual of
the $j$-th (not converged) Ritz pair is maximized, while the 
number of floating-point operations (FPOs) is minimized. 
In other words, at each restart, we try to find the optimal 
$ k $ by maximizing the object function \cite{nu-TRLan}
\beq
f(k) = \frac{d_j}{\mathrm{\#\ of\ FPOs}}, \hspace{5mm}
d_j \simeq \mathcal{C}_{m-k}(1+2\gamma), \hspace{5mm}
\gamma = \frac{\hat\lambda_{k+1} - \hat\lambda_{j}}
              {\hat\lambda_{m} - \hat\lambda_{k+1}}, 
\eeq
where $\mathcal{C}_{m-k}(z)$ is the Chebyshev polynomial of degree $m-k$.
For the number of FPOs, we only count the dominant parts which are directly
related to the dimension $m$, namely, the re-orthogonalization and 
the update of Ritz vectors.
This modified algorithm is called the {\em Adaptive Thick-Restart Lanczos algorithm}, 
which can attain substantial performance gain, with faster convergence
and fewer FPOs. 


\begin{figure}[htb]
\begin{tabular}{@{}c@{}c@{}}
\includegraphics[width=75mm,clip=true]{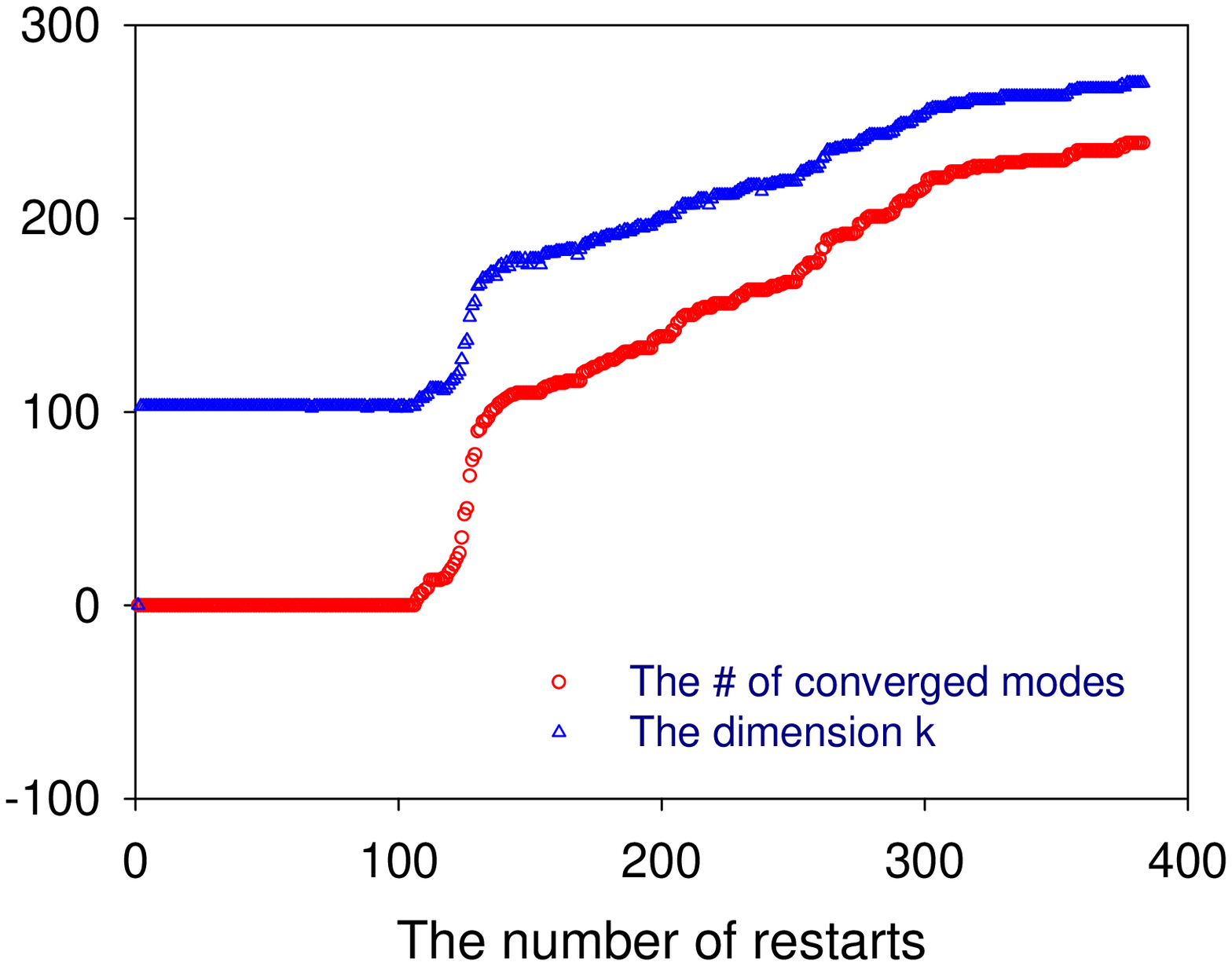} &
\includegraphics[width=75mm,clip=true]{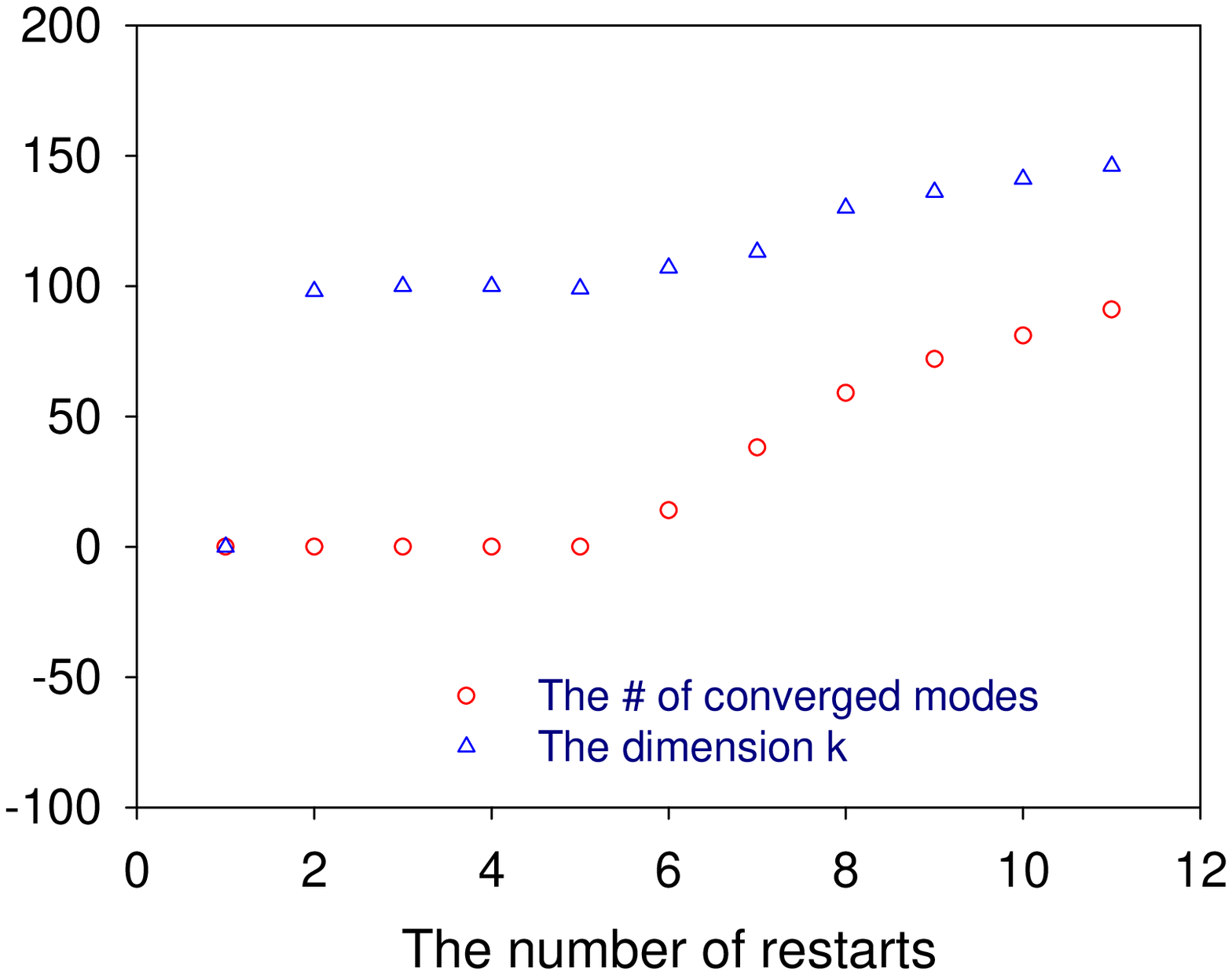} \\
(a) & (b)
\end{tabular}
\caption{\label{fig:nuTRLan}
The number of converged Ritz pairs and the dimension $ k $
versus the number of restarts for the $\nu$-TRLan algorithm: 
(a) $H_w$ ($m=340$); 
(b) $ S_+ $ of the overlap operator ($m=240$).}
\end{figure}

For benchmarking, we project the low-lying eigenmodes
of $H_w$ and $S_+$, using a nontrivial configuration ($ Q_t = 3 $) 
generated by the 2-flavor QCD simulation on a $16^3 \times 32$ lattice, 
with ODWF fermion at $N_s=16$ and $ m_{sea} a = 0.01 $, 
and the plaquette gauge action at $\beta=5.90$. 
In Table \ref{tab:benchmark}, we present our benchmark results 
for three different methods: 
ARPACK (Arnoldi algorithm with Implicit Restart), 
TRLan (Thick-Restart Lanczos), 
and $\nu$-TRLan (Adaptive Thick-Restart Lanczos).
For TRLan (by definition $k=k'$, where 
$k'$ is the number of low-modes we want to project,  
and $k$ is the truncated dimension at each restart),  
we see that its rate of convergence tends to be very slow for $H_w$,
with a rather large condition number ($\sim 10^6$).
On the other hand, for $\nu$-TRLan, $k$ is automatically adjusted to be an optimal
value in each restart ($k' < k < m$, see Fig. \ref{fig:nuTRLan}),
thus the rate of convergence is dramatically improved. It turns out that,
for both $H_w$ and $S_\pm $, $\nu$-TRLan is faster than
TRLan and ARPACK. Therefore, we use $\nu$-TRLan to project the low-lying 
eigenmodes of $H_w$ and $D$.

\section{Lattice setup}

Simulations are carried out for two flavors ($N_f=2$) QCD on a $16^3 \times 32$
lattice at a lattice spacing $a \sim $ 0.11~fm, for eight sea quark masses in the range
$ m_q a =0.01, 0.02, \cdots, 0.08 $ \cite{Chiu:2009wh}.
For the gluon part, the plaquette action is used at $\beta$ = 5.90.
For the quark part, the optimal domain-wall fermion is used with $ N_s = 16 $. 
After discarding 300 trajectories for thermalization, we accumulated
$ 3000-3200 $ trajectories in total for each sea quark mass.
From the saturation of the binning error of the plaquette, as well as 
the evolution of the topological charge, we estimate the autocorrelation time 
to be around 10 trajectories. Thus we sample one configuration every 
10 trajectories, then we have $ 270-290 $ configurations for each sea quark mass.

For each configuration, we calculate the zero modes and 
80 conjugate pairs of the lowest-lying eignmodes of the overlap Dirac operator.
We outline our procedures as follows.
First, we project 240 low-lying eigenmodes of $ H_w^2 $ using $\nu$-TRLan 
alogorithm, where each eigenmode has a residual less than $ 10^{-12} $.
Then we approximate the sign function of the overlap operator  
by the Zolotarev optimal rational approximation with 64 poles,   
where the coefficents are fixed with $ \lambda_{max}^2 = (6.4)^2 $,  
and $ \lambda_{min}^2 $ equal to the maximum of 
the 240 projected eigenvalues of $ H_w^2 $. 
Then the sign function error is less than $ 10^{-14} $.
Using the 240 low-modes of $ H_w^2 $ and the Zolotarev approximation 
with 64 poles, we project the zero modes plus 80 conjugate pairs of 
of the lowest-lying eignmodes of the overlap operator 
with the $\nu$-TRLan algorithm, 
where each eigenmode has a residual less than $ 10^{-12} $.

\section{Results}

\begin{figure}[htb]
\centerline{\includegraphics[width=100mm,clip=true]{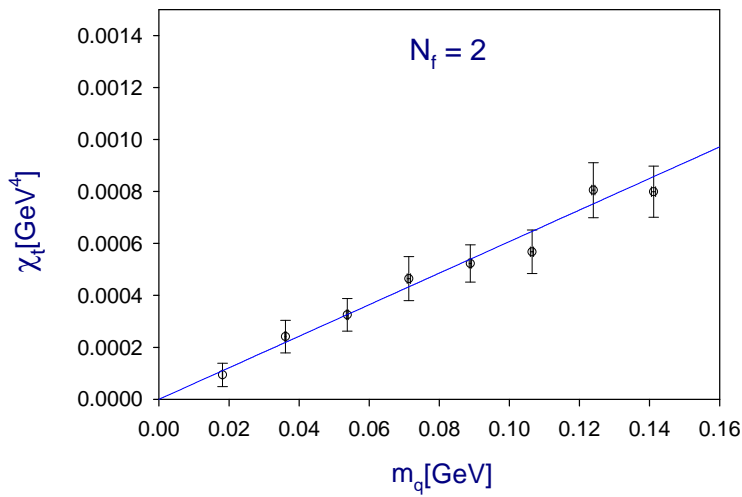}}
\caption{\label{fig:chit_mq_nf2}
The topological susceptibility of two-flavors QCD with the optimal domain-wall fermion.}
\end{figure}

In Fig.~\ref{fig:chit_mq_nf2}, we plot the topological susceptibility
$ \chi_t = \left< (Q_t-\left<Q_t\right>)^2 \right> / \Omega $ 
as a function of the sea quark mass $m_q $. 
For all eight sea quark masses, our data are
well fitted by a linear function $F+G m_q$ with the intercept
$F = 0.47(2.37) \times 10^{-5} $ and the slope $G = 5.50(4) \times 10^{-3}$.
Evidently, the intercept is consistent with zero, in agreement
with the $\chi$PT expectation (\ref{eq:chi_ChPT}).
Equating the slope to $ \Sigma/N_f$, we obtain
$ \Sigma = 0.0110(7)$.
In order to convert $\Sigma$ to that in the
$\overline{\mathrm{MS}}$ scheme, we calculate the
renormalization factor $Z_m^{\overline{\mathrm{MS}}}(\mathrm{2~GeV})$
using the non-perturbative renormalization technique
through the RI/MOM scheme \cite{Martinelli:1994ty}, and
obtain $Z_m^{\overline{{\mathrm{MS}}}}(\mbox{2~GeV}) = 0.5934(10)$ \cite{Chiu:NPR}.
Then the value of $ \Sigma $ is transcribed to
\bea
\label{eq:sigmaMS}
\Sigma^{\overline{{\mathrm{MS}}}}(\mathrm{2~GeV}) = [\mathrm{261(5)(7)~MeV}]^3, 
\eea
which is in agreement with our previous results : 
$ \Sigma^{\overline{\mathrm{MS}}}(\mathrm{2~GeV})
= [\mathrm{245(5)(12)~MeV}]^3 $
for $ N_f = 2 $ lattice QCD with overlap fermion 
in a fixed topology \cite{Aoki:2007pw},  
and 
$ \Sigma^{\overline{\mathrm{MS}}}(\mathrm{2~GeV})
= [\mathrm{253(4)(6)~MeV}]^3 $ 
for $ N_f = 2+1 $ lattice QCD with domain-wall fermion \cite{Chiu:2008jq}. 
The statistical error represents a combined error
(including those due to $a^{-1}$ and $Z_m^{\overline{\mathrm{MS}}}$).
The systematic error is estimated from the turncation of higher order 
effects and the uncertainty in the determination of lattice spacing with 
$ r_0 = 0.49 $ fm.
Since our calculation is done at a single lattice spacing,
the discretization error cannot be quantified reliably, but
we do not expect much larger error because our lattice
action is free from $O(a)$ discretization effects.


\section{Concluding remarks}

We have measured the topological charge
of the gauge configurations generated by lattice simulations
of 2 flavors QCD on a $ 16^3 \times 32 $ lattice,  
with Optimal Domain-Wall Fermion (ODWF) at $ N_s = 16 $  
and plaquette gauge action at $ \beta = 5.90 $.  
We use the Adaptive Thick-Restart Lanczos algorithm to compute the low-lying eigenmodes
of $H_w^2$ and the overlap Dirac operator.
Our result of $\chi_t$ agrees with 
the sea-quark mass dependence predicted by the ChPT at the tree level, 
and provides a good determination of the chiral condensate.
Furthermore, our result of $ \chi_t $ also agrees with the 
the NLO ChPT, Eq. (\ref{eq:chi_ChPT_NLO}).
We will present the details in a forthcoming paper \cite{Chiu:2010}.

\begin{acknowledgments}
  This work is supported in part by the National Science Council
  (Nos.~NSC96-2112-M-002-020-MY3, NSC99-2112-M-002-012-MY3, NSC96-2112-M-001-017-MY3, 
  NSC99-2112-M-001-014-MY3, NSC99-2119-M-002-001) and NTU-CQSE~(Nos.~99R80869,~99R80873).
  We also thank NCHC and NTU-CC for providing facilities to perform some of the computations. 
\end{acknowledgments}


\end{document}